\documentclass[useAMS]{mn2e}
\usepackage{graphicx}	
\usepackage{psfig}
 \usepackage{url}
\usepackage{times}
\def\spose#1{\hbox to 0pt{#1\hss}}
\newcommand\lsim{\mathrel{\spose{\lower 3pt\hbox{$\mathchar"218$}}
     \raise 2.0pt\hbox{$\mathchar"13C$}}}
\newcommand\gsim{\mathrel{\spose{\lower 3pt\hbox{$\mathchar"218$}}
     \raise 2.0pt\hbox{$\mathchar"13E$}}}
\def\sc{Schwarzschild}
\def\ltsima{$\; \buildrel < \over \sim \;$}
\def\lsim{\lower.5ex\hbox{\ltsima}}
\def\gtsima{$\; \buildrel > \over \sim \;$}
\def\gsim{\lower.5ex\hbox{\gtsima}}

\title[The Fermi blazar sequence]
{The {\it Fermi} blazar sequence}
\author[G. Ghisellini et al.]
{G. Ghisellini$^1$ \thanks{E--mail: gabriele.ghisellini@brera.inaf.it},
C. Righi$^{2,1}$, L. Costamante$^1$, F. Tavecchio$^1$\\ \\
$^1$ INAF -- Osservatorio Astronomico di Brera, via E. Bianchi 46, I--23807 Merate, Italy \\
$^2$ DiSAT, Universit\`a degli Studi dell'Insubria, Via Valleggio 11, I--22100 Como, Italy \\
}

\begin{document}

\pagerange{\pageref{firstpage}--\pageref{lastpage}} \pubyear{2012}

\maketitle
\label{firstpage}

\begin{abstract}
We revisit the blazar sequence exploiting the complete, flux limited sample
of blazars with known redshift detected by the {\it Fermi} satellite after 4 years of operations
(the 3LAC sample).
We divide the sources into $\gamma$--ray luminosity bins, collect all the archival data for all 
blazars, and construct their spectral energy distribution (SED).
We describe the average SED of blazars in the same luminosity bin through a  simple
phenomenological function consisting of two broken power laws connecting with
a power law describing the radio emission.
We do that separately for BL Lacs and for flat spectrum radio quasars (FSRQs) and also for
all blazars together. 
The main results are:
i) FSRQs display approximately the same SED as the luminosity increases,  
but the relative importance of the high energy peak increases;
ii) as a consequence, X--ray spectra of FSRQs become  harder for larger luminosities;
iii) BL Lacs form indeed a sequence: they become redder (i.e. smaller peak frequencies) 
with increasing luminosities, with a softer  $\gamma$--ray slope and a larger 
dominance of the high energy peak;
iv) for all blazars (BL Lacs+FSRQs) these properties becomes more prominent, as the highest
luminosity bin is populated mostly by FSRQs and the lowest luminosity bin mostly by BL Lacs.
This agrees with the original blazar sequence, although BL Lacs never have an average $\gamma$--ray
slope as hard as found in the original sequence.
v) At high luminosities, a large fraction of FSRQs shows signs of thermal emission from the accretion
disc, contributing in the optical--UV. 
\end{abstract}

\section{Introduction}

Flat Spectrum Radio Quasars (FSRQs) and BL Lac objets form the class of blazars, 
namely jetted Active Galactic Nuclei whose jet is pointing towards us.
Since the emitting plasma in the jet moves at relativistic speeds, the 
emission is strongly beamed, making blazars the most powerful 
persistent objects in the Universe, and visible also at large redshifts $z$.
The operative difference between the two flavours of blazars concerns the
importance of the broad emission lines with respect to the underlying continuum,
once the blazar nature is confirmed (strong radio emission with respect to the optical,
strong X--ray emission, possibly strong $\gamma$--ray emission).
If the (rest frame, when $z$ is known) equivalent width EW of the lines is larger than 5 \AA\ 
the source is a FSRQ, if the EW$<5$\AA\ the source is a BL Lac (Urry \& Padovani 1995).

This classification criterion is easy and practical, especially when no lines
is visible, but does not always correspond to a physical distinction.
An alternative way, proposed by Ghisellini et al. (2011), is to measure the
luminosity $L_{\rm BLR}$ of all broad lines in units of the Eddington luminosity
($L_{\rm EDD}$), and call FSRQs the sources whose $L_{\rm BLR}\gsim 10^{-3} L_{\rm Edd}$,
and BL Lacs the others. 
This is due to the change of the regime of accretion supposed to occur at 
disc luminosities $L_{\rm disc}/L_{\rm Edd}\sim 10^{-2}$, and to the
fact that, on average, $L_{\rm BLR}\approx 0.1 L_{\rm disc}$, although with a large  
dispersion (see e.g. \S 2.2 in Calderone et al. 2013).
Below this value the disc is thought to become radiatively inefficient, its
UV ionizing luminosity is a minor fraction of its bolometric output 
and it is not enough to photo--ionize the broad line clouds
(e.g. Sbarrato, Padovani \& Ghisellini 2014).
The disadvantage of this, more physical, classification scheme is that, besides the
redshift, the black hole mass must be known.

The overall spectral energy distribution (SED) of blazars consists of 
two broad humps, peaking in the IR--X--ray band and in the MeV--TeV band,
and,  for the most powerful blazars, there is often a 
third peak due to the contribution of the accretion disc.
Quasi--simultaneous variability in different spectral bands is often (but not always) 
observed, suggesting that most of the flux is produced in a single region of the jet.
This does not exclude that different zones at different scales of the jet
can contribute, at least to dilute the variability amplitude of the main component.
After all, we do see, in VLBI radio maps, distinct blobs at different distances
from the central black hole.
These regions are mainly responsible for the radio flux observed at frequencies
smaller than $\sim$100 GHz, characterized by variability timescales much longer
than what observed from the optical to the $\gamma$--rays.
The flux in the latter bands must be produced in compact regions, presumably  much
closer to the black hole: at these scales the radio emission is self--absorbed.

\subsection{The original blazar sequence}

\begin{figure} 
\vskip -0.6 cm
\hskip -0.1 cm
\includegraphics[height=9.2cm]{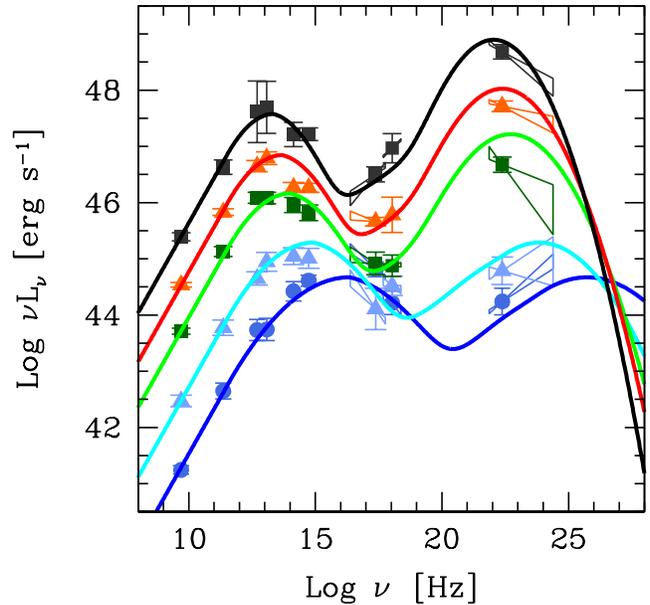}
\caption{ 
The original blazar sequence (Fossati et al. 1998; Donato et al. 2001),
constructed with the 126 blazars belonging to complete 
(flux limited) radio and X--ray samples of blazars at those times. 
Only 33 out of the 126 blazars were detected by {\it CGRO}/EGRET.
For the lowest luminosity bin, only 3 blazars were detected by EGRET.
} 
\label{seq1}
\end{figure}

The number of blazars considered for the original blazar sequence 
by Fossati et al. (1998, hereafter F98) was 126.
These objects belonged to the existing (at the time) complete (flux limited) samples,
namely one X--ray  and two radio
samples (Elvis et al. 1992; K\"uhr et al. 1981; Padovani \& Urry 1992; Wall \& Peacock 1985).
Only 33 were detected in the $\gamma$--ray band by
the EGRET instrument onboard the {\it Compton Gamma Ray Observatory}.
Given the limited sensitivity of EGRET, these 33 blazars were the 
brightest $\gamma$--ray blazars at that time.
The 126 objects were divided into 5 GHz radio luminosity bins,
their luminosities at selected frequencies were averaged,
in order to build the average SED of blazars in each bin of radio luminosity.
For the lowest radio luminosity bin, only 3 blazars were detected by EGRET:
Mnk 421, PKS 2005--489 and PKS 2155--304. 
Donato et al. (2001) later considered the slope of the X--ray emission for the
same objects, that was added to the average SEDs.
The result is shown by Fig. \ref{seq1}.
The main result of this study was that the shape of the SED 
changed in a coherent way as the radio luminosity changes.
Since the latter is a good tracer of the bolometric luminosity $L_{\rm bol}$,
we can equivalently say that the sequence in Fig. \ref{seq1} shows trends 
controlled by one parameter, i.e. the total {\it observed} luminosity:
\begin{enumerate}
\item blazars become ``redder" with increasing $L_{\rm bol}$, namely the
peak frequencies become smaller;
\item at the same time, the ``Compton dominance" increases, namely
the ratio of the luminosity of the high energy hump (that is interpreted as due to
the inverse Compton process, hence the name) over the low energy hump
(due to synchrotron) increases;
\item the $\gamma$--ray slope become softer with increasing $L_{\rm bol}$;
\item the X--ray slope becomes harder with increasing $L_{\rm bol}$.
\end{enumerate}

Ghisellini et al. (1998) soon interpreted these properties as due to the different
amount of radiative cooling suffered by the emitting electrons in different sources, 
implicitly assuming that the heating mechanism, instead, was similar for all.
For this explanation, it is crucial to notice that 
the broad line clouds in FSRQs (that are more luminous than BL Lacs)
can re--isotropize $\sim$10\% of the disc radiation.
An even larger fraction can be re--isotropized and re--emitted in the infrared band 
by the absorbing torus surrounding the disc.
These radiation components are seen relativistically boosted in the frame 
comoving with the emitting region of the jet, but they are present only in 
FSRQs, not in BL Lacs. 
Therefore in FSRQs the inverse Compton luminosity is large, due to the presence
of external seed photons, and this explains a larger Compton dominance
(Sikora, Begelman \& Rees 1994).
Furthermore, since the cooling rate is more severe, electrons attain
smaller typical energies, explaining why FSRQs have ``redder" SED.

In BL Lacs, instead, the lines and the torus (see Chiaberge et al. 1999) are absent. 
The seed photons for the IC process are due to the internally
produced synchrotron radiation, and this implies larger typical
electron energies (they suffer less radiative cooling) and a smaller
Compton dominance.
The other pillar for explaining the sequence is that lineless 
BL Lacs are less luminous than FSRQs.
At the low luminosity extreme,
the bluest BL Lacs should not be strong MeV or GeV emitters and
can even be missed by {\it Fermi}/LAT (see e.g. Bonnoli et al. 2015).

The fact that the {\it observed} bolometric luminosity is the parameter
governing the original blazar sequence is, at first sight, strange.
In fact the observed luminosity is the received flux, assumed isotropic,
multiplied by $4\pi d_{\rm L}^2$, where $d_{\rm L}$ is the luminosity distance.
But the emission is instead strongly boosted, so the observed flux depends on 
the viewing angle $\theta_{\rm v}$ and the bulk Lorentz factor $\Gamma$,
that form the beaming factor $\delta=1/[\Gamma (1-\beta\cos\theta_{\rm v})]$.
Can the sequence be a sequence of $\delta$, of objects that instead share the same
intrinsic (i.e. in the comoving frame) properties (as proposed by Nieppola et al. 2008)?
In the  study of Nieppola et al. (2008) the $\delta$--factor  is derived through variability of the 
radio flux, yielding a lower limit on the brightness temperature,
and a corresponding lower limit to $\delta$.
On the other hand, direct VLBI and VLBA measurements of superluminal knots in blazars
(Jorstad et al. 2001; Lister et al. 2013), and modelling (Ghisellini \& Tavecchio 2015)
do not find strong differences in the bulk Lorentz factors of BL Lacs and FSRQs.

In recent years a correlation between the jet power and the accretion luminosity
has been found, with some evidence that the jet power is larger than the disc luminosity
(e.g. Ghisellini et al. 2014).
Thus we can wonder if the blazar sequence is mainly due to a sequence of 
accretion rates.
We believe that this is certainly the case, but at two conditions: 
i) the viewing angle $\theta_{\rm v}$ is small for all objects (of the
order of $1/\Gamma$, to ensure maximal Doppler boost), and
ii) the black hole mass of all blazars are similar (and large), 
so that the accretion rate, in absolute terms, corresponds to the accretion 
rate in Eddington units.
In fact, as mentioned above, the accretion regime changes when the disc
luminosity is of the order of $10^{-2}L_{\rm Edd}$ (and this should be the cause
of the ``blazars' divide", Ghisellini, Maraschi \& Tavecchio 2009).
If this is a sharp threshold or instead there is a smooth transition
is a matter of debate (eg. Narayan et al. 1997, Sharma et al. 2007), but 
the existence of a transition between a radiatively efficient and a radiatively 
inefficient regime is rather certain. 
This would explain the absence of strong broad emission lines in BL Lacs,
and their different SED.

Is the hypothesis of ``same black hole mass for all blazars" reasonable?
We believe that for the original sequence there is a strong selection effect making
this hypothesis true: given the sensitivity limits of the instruments of $\sim$20
years ago (especially in the $\gamma$--ray band), only the most powerful objects could be 
observed, and therefore the ones with the largest black hole masses. 
For BL Lacs we have the following chain of arguments:
even if BL Lacs are not very powerful, the emission of their jet is observed
$\to$ the  jet power is larger than some limit 
$\to$ the accretion rate is larger than some limit $\to$ the black
hole mass must be large in order to have a radiatively inefficient disc
(i.e. the luminosity is much smaller than the Eddington one, yet observable).
For powerful FSRQs we have:
the jet emission is very powerful $\to$ the jet power is large $\to$
the accretion rate is large {\it and} the black hole mass is large
(i.e. having luminosities close to Eddington but small black hole masses
correspond to jets not luminous enough to be observed).

One of the motivation of the present work is to check what we can infer now
that we have more sensitive instruments: {\it Fermi}/LAT is $\sim$20 times 
more sensitive than EGRET, and patrols the entire sky in a much more efficient way.
The expectation is that we will start to see blazars with black hole masses
smaller than the ones forming the original blazar sequence.
In this case we ought to see a FSRQ with a smaller bolometric observed luminosity 
(similar to a BL Lac belonging to the original sequence), but with ``red" SED very 
typical of FSRQs.
This is because we expect that the parameter controlling the sequence is not the 
mere luminosity, but the luminosity in units of Eddington,
as proposed by Ghisellini \& Tavecchio (2008).

Using the same arguments we expect  that an improved sensitivity 
allows to detect also slightly misaligned objects, namely
``blazars at the edge" with $\theta_{\rm v} \gsim 1/\Gamma$.
Since the emission of these objects is less beamed and less blueshifted, we
should observe FSRQs with a small observed bolometric luminosity that
are (slightly) ``redder" that their powerful aligned cousins and much redder than
BL Lacs with the same bolometric luminosity.

Despite the success to correctly describe the {\it observed} blazars 
and to explain them in a simple theoretical scheme, 
the existence of the blazar sequence was and still is a very debated issue.
The main objection is that it could be the result of selection effects that are
operating not only when it was proposed, but even now, despite the presence of
more sensitive instruments and more extended complete sample down to deeper
sensitivity limits
(Giommi, Menna \& Padovani,	1999;
Perlman et al. 2001;
Padovani et al. 2003;
Caccianiga \& Marcha 2004;	
Ant\'on \& Browne 2005;
Giommi et al. 2005;
Nieppola, Tornikoski \& Valtaoja 2006;
Raiteri \& Capetti (2016);
Padovani, Giommi \& Rau 2012; see also the reviews by Padovani 2007 and Ghisellini \& Tavecchio 2008).

To demonstrate the point, it has been recently suggested an alternative scheme,
the  ``simplified blazar scenario" (Giommi et al. 2012).
There are two crucial assumptions in this scheme:
i) a pre--assigned, given distribution of electron energies responsible for the spectral peaks of the SED
(or random Lorentz factor $\gamma_{\rm peak}$),  
ii) $\gamma_{\rm peak}$ does not correlate with the observed jet luminosity $L$.
This contrasts with the blazar sequence view, in which $\gamma_{\rm peak}$ inversely
correlates with $L$. 
In the ``simplified blazar scenario" we can have blue and red blazars at all luminosities.
Then, through simple prescriptions (e.g. all blazars have the same magnetic field and Doppler factor),
but taking into account current sensitivity limits in different frequency bands, 
Giommi et al. (2012) can reproduce the existing observations.
This scheme therefore, does not question the reality of the {\it observed} sequence, but
interprets it as the outcome of selection effects.
This is done at the cost of introducing a given $\gamma_{\rm peak}$ distribution,
which has no physical explanation yet, and abandoning the idea of the different radiative cooling
operating in different blazars (see also \S \ref{simplified}).

Bearing in mind all the above, we feel motivated to revisit the blazar sequence scenario.
In the first 4 years of operations, {\it Fermi} detected about 1500 blazars and for 
nearly one half of them we know the redshift (Ackermann et al. 2015).
Besides the impressive amount of $\gamma$--ray data, we can now take advantage
of the SDSS survey and the {\it Planck}, {\it WISE} and {\it Swift} satellites.
Since the $\gamma$--ray luminosity is often the dominant part of the electromagnetic
output, and given the abundance of $\gamma$--ray sources, we use, as the only criterion for
selection, the detection by {\it Fermi} and the knowledge of the redshift.
This is different from the original blazar sequence, where the objects were selected on the
basis of their radio and/or X--ray emission.

During the completion of this work, also Mao et al. (2016) revisited the
blazar sequence by considering a very large number of blazars
(above two thousands) included in the Roma--BZCAT catalog.
They find that the general trends of the original F98 sequence 
is confirmed, but with a large scatter for the ensemble of
blazars, without dividing them into BL Lacs and FSRQs.

We use a flat cosmology with $h=\Omega_\Lambda=0.7$.

\begin{figure} 
\vskip -0.5 cm
\hskip -0.2 cm
\includegraphics[height=9.5cm]{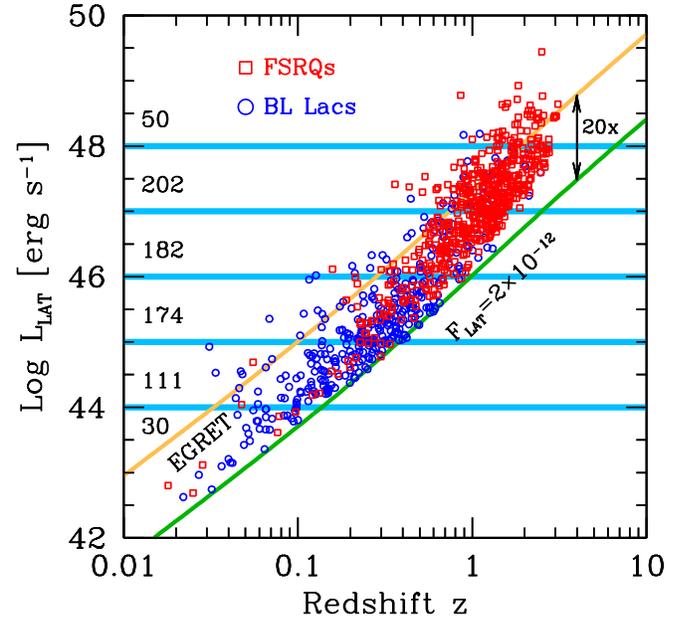}
\vskip -0.5 cm
\caption{
The K--corrected $\gamma$--ray luminosity (in the rest frame 0.1--100 GeV band) 
as a function of redshift for all blazars in the 3LAC catalog with known redshift.
The solid lines refer to approximately the sensitivity limit of EGRET (onboard the {\it CGRO} 
satellite (orange line) and the {\it Fermi}/LAT sensitivity of the 3LAC sample,
that is $\sim$20 times deeper.
The horizontal lines divide the sample into 6 $\gamma$--ray luminosity bins,
and the corresponding total number of sources in each bin is indicated.
Red squares refer to FSRQs, blue circles to BL Lacs, following the classification 
of the 3LAC catalog.
} 
\label{lgz}
\end{figure}
\begin{figure} 
\vskip -0.5 cm
\hskip -0.5 cm
\includegraphics[height=9.5cm]{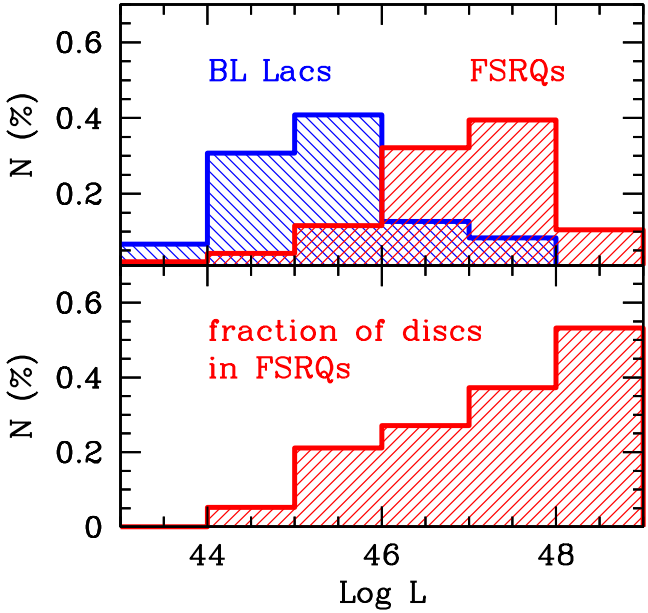}
\vskip -0.5 cm
\caption{Top: Fraction of FSRQs and BL Lacs as a function of redshift.
Bottom: fraction of FSRQs showing evidence of accretion disc emission
in their SED, besides the presence of broad emission lines.
} 
\label{bbb}
\end{figure}

\section{The sample}

We consider the third catalog of AGN detected by {\it Fermi}/LAT 
(3LAC; Ackermann et al. 2015).
This catalog lists the $\gamma$--ray luminosity  averaged over 4 years of {\it Fermi}/LAT observations.
{\it Fermi}/LAT patrols the entire sky in 3 hours, and  its sky sensitivity map is rather uniform
over the entire sky.
The 3LAC sample can then be considered as a complete, flux limited sample.
In includes all the detected AGNs detected between 100 MeV and 300 GeV
with a Test Statistic (Mattox et al. 1996) greater than 25 (roughly equivalent to 5$\sigma$).
There are 1563 sources identified with high confidence with AGNs 
at high Galactic latitudes ($\vert b\vert >10^\circ$), 
and 98\% of these are blazars. 
About half of the newly detected blazars are of unknown type, i.e., they 
lack spectroscopic information of sufficient quality to determine the strength 
of their emission lines. 
Based on their $\gamma$--ray spectral properties, these sources 
are split between flat--spectrum radio quasars (FSRQs) and BL Lacs
(following the division in Ghisellini, Maraschi \& Tavecchio  2009).
About 50\% of the BL Lacs have no measured redshifts. 
Sources with no multiple identifications and with no analysis issues 
(see \S2 in Ackermann et al. 2015), form the ``clean" sample that includes 
414 FSRQs, 604 BL Lacs, 402 blazars of unknown type and 24 non--blazar AGNs.
Out of the blazars of unknown type, 40 sources are in the clean sample with 
spectroscopic redshift, but with spectra of low quality (bcu I type in
Ackermann et al. 2015).
Out of this sample, 34 blazars with redshift have been detected in the TeV band.
We consider the blazars with redshift contained in the ``clean" 3LAC catalog.
Excluding the objects classified as non blazars AGNs or Narrow Line Seyfert Galaxies, 
we select 747 objects classified as BL Lacs (299) or FSRQs (448).
For each of them we constructed the overall SED, using the ASI Astrophysical Data Center (ASDC) 
database\footnote{http://www.asdc.asi.it/fermi3lac/}.
We calculate the K--corrected $\gamma$--ray luminosity in the 0.1--100 GeV range,
using the $\gamma$--ray spectral index listed in the 3LAC catalog. 
Fig. \ref{lgz} shows the {\it average} (over 4 years) $\gamma$--ray luminosity 
as a function of redshift. 
Blue circles are BL Lacs, red squares are FSRQs, as defined by Ackermann et al. (2015).
We show also the line corresponding to a
flux limit of $2\times 10^{-12}$ erg cm$^{-2}$ s$^{-1}$ in the 0.1--100 GeV band
(appropriate for {\it Fermi}/LAT), and
a line corresponding to 20 times this value, to mimic the approximate limit of EGRET.
It is clear that  BL Lacs have lower redshifts and smaller $\gamma$--ray luminosities
(K--corrected, integrated in the rest frame 0.1--100 GeV band) than FSRQs.
The fraction of BL Lacs and of FSRQs as a function of 
$L_\gamma$ is shown in the top panel of Fig. \ref{bbb}.
 
We divided these blazars into 6 $\gamma$--ray luminosity bins.
The number of sources in each bin is reported in Fig. \ref{lgz}.
Each bin spans a decade in $\gamma$--ray luminosity, 
for an easy comparison with the old sequence, whose radio luminosity
bins were also a decade.
We also tried different binning, with very similar results.

\begin{figure} 
\vskip -0.4 cm
\hskip -0.5 cm
\includegraphics[height=9.5cm]{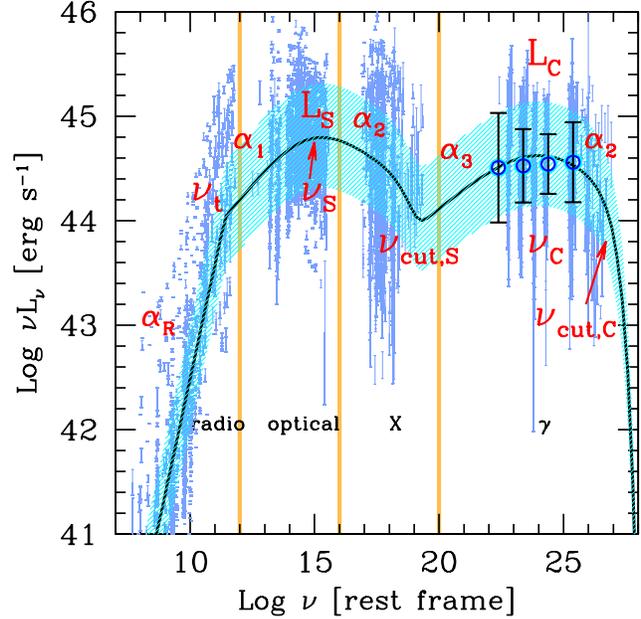}
\vskip -0.5 cm
\caption{Sketch illustrating the phenomenological description of the 
average SED of the blazars in each luminosity bin.
In this example we show the SED of the BL Lacs in the bin 
$45 <\log (L_\gamma/{\rm erg\, s^-1})<46$.
We show the data of each source, with no averaging.
The hatched stripe corresponds to 1$\sigma$ of the dispersion of points
around the fitting law showed with the black solid line.
The red labels correspond to the needed parameters.
} 
\label{sketch}
\end{figure}

\begin{table*} 
\centering
\begin{tabular}{lllll lllll llll }
\hline
\hline
$\log L_\gamma$ &$\alpha_1$ &$\alpha_2$ &$\alpha_3$ &$\nu_{\rm t}$ &$\nu_{\rm S}$ &$\nu_{\rm C}$ &$\nu_{\rm cut, S}$
&$\nu_{\rm cut, C}$ &$\nu_{\rm S}L(\nu_{\rm S})$ &CD   &$\sigma$ &N\\
$\log$(erg s$^{-1}$)    &           &            &        &Hz       &Hz      &Hz    &Hz  &Hz  &erg s$^{-1}$ & & & \\
\hline   
 & & & & & {\bf All} & & & & &\\
 \hline
$>48$   &0.5  &1.45 &0.1  &1e12 &2.5e12 &9e20 &2e16 &3e26   &1.3e47 &15   & --0.31, 0.57 &49  \\
47--48  &0.5  &1.42 &0.24 &1e12 &2.5e12 &2e21 &1e16 &3e27   &5e46   &4.8  & --0.33, 0.52 &202   \\
46--47  &0.7  &1.38 &0.5  &6e11 &5e12   &2e21 &1e19 &6e26   &1e46   &2    & --0.49, 0.42 &182 \\
45--46  &0.7  &1.15 &0.65 &4e11 &5e12   &1e22 &1e19 &6e27   &7e44   &0.6  & --0.54, 0.51 &174  \\
44--45  &0.7  &1.15 &0.86 &1e11 &1e16   &8e24 &1e19 &1e27   &1.5e44 &0.35 & --0.83, 0.86 &111 \\
$<44$   &0.75 &1.3  &0.93 &9e10 &4e16   &3e25 &1e19 &2e27   &2e43   &0.25 & --0.68, 1.09 &29   \\

\hline   
 & & & & & {\bf FSRQs} & & & & &\\
 \hline
$>48$   &0.5  &1.45 &0.1  &1e12 &1e12   &9e20 &2e16 &3e26   &1.3e47  &15   & --0.43, 0.71 &47 \\
47--48  &0.5  &1.45 &0.22 &1e12 &1e12   &1e21 &1e16 &3e27   &5e46    &5.3  & --0.37, 0.42 &177 \\
46--47  &0.5  &1.38 &0.4  &8e11 &2e12   &1e21 &7e15 &6e26   &1e46    &2.4  & --0.33, 0.61 &144 \\
45--46  &0.5  &1.42 &0.55 &5e11 &4.5e12 &2e21 &1e16 &3e26   &1.5e45  &1.7  & --0.32, 0.55 &52  \\
44--45  &0.5  &1.33 &0.75 &3e11 &6e12   &5e21 &1e16 &6e26   &2e44    &0.7  & --0.66, 0.84 &19  \\ 
$<44^*$ &---  &---  &---  &---  &---    &---  &---  &---    &---     &---  & &9  \\
\hline   
 & & & & &{\bf BL Lacs} & & & & &\\
 \hline
$>48^*$ &---  &---  &---  &---    &---  &---    &---  &---  &---    &---  & &2  \\
47--48  &0.5  &1.2  &0.45 &7e11   &8e11 &1.e21  &5e15 &3e26 &2.5e46 &2.7  & --0.31, 0.49 &25 \\
46--47  &0.65 &1.13 &0.62 &5e11   &5e12 &1.5e21 &6e15 &7e26 &4e45   &1.2  & --0.44, 0.45 &38 \\
45--46  &0.7  &1.2  &0.8  &2.5e11 &8e14 &1e24   &6e18 &1e27 &6e44   &0.7  & --0.48, 0.52 &122 \\
44--45  &0.68 &1.2  &0.8  &1e11   &1e16 &8e24   &4e19 &8e27 &1.5e44 &0.4  & --0.69, 0.74 &92  \\ 
$<44$   &0.72 &1.15 &0.87 &8e10   &8e16 &3e25   &5e19 &3e28 &2.5e43 &0.28 & --0.48, 0.75 &20 \\
\hline
\hline
\end{tabular}
\caption{Parameters used to the phenomenological SEDs 
shown in Fig. 1.
For all models we kept fixed $\alpha_{\rm R}=-0.1$.
$^*$: in these luminosity bins, the number of objects (last column)
is too small for a meaningful derivation of mean values.
For this reason, we do not show the corresponding curve in Fig. \ref{seq2}.
}
\label{para}
\end{table*}

\section{Phenomenological SEDs}

The entire non--thermal SED of all blazars can be described by
two broad humps and a flat radio spectrum.
The simplest analytical function to approximate such
a broad band SED is a single power in the radio, connecting
to a smoothly broken power law, describing the low energy hump,
plus another smoothly broken power law that describes the high energy hump.
The most remarkable  deviations from this simple description are related
either to the emission from the accretion disc and the torus (in the optical--UV
and the far IR, respectively) and the emission of the host galaxy visible at low
redshifts.
Therefore we propose to describe the entire non--thermal SED 
with the following prescription, completely phenomenological.
In the radio band we have
\begin{equation}
L_{\rm R}(\nu) \, =\, A\, \nu^{-\alpha_{\rm R}}; \quad \nu \le\nu_{\rm t}
\label{eq1}
\end{equation}
where $\nu_{\rm t}$ is where the flat part ends.
It can be interpreted as the self--absorption frequency of the most
compact jet emitting region. 
This power law connects with:
\begin{equation}
L_{\rm S+C}(\nu) \, = L_{\rm S}(\nu)+L_{\rm C}(\nu); \, \, \nu > \nu_{\rm t}
\label{eq2}
\end{equation}
describing the low and the high frequency part, respectively.
The low energy part (that can be associated to synchrotron flux) is
assumed to be:
\begin{equation}
L_{\rm S}(\nu) \, =\, B\, { (\nu/\nu_{\rm S})^{-\alpha_1}
\over 1+(\nu/\nu_{\rm S})^{-\alpha_1+\alpha_2}} \, \exp(-\nu/\nu_{\rm cut, S}); \, \, \nu > \nu_{\rm t}
\end{equation}
while the high energy part (that can be associated to the inverse Compton flux) is:
\begin{equation}
L_{\rm C}(\nu) \, =\, C\, { (\nu/\nu_{\rm C})^{-\alpha_3}
\over 1+(\nu/\nu_{\rm C})^{-\alpha_3+\alpha_2}} \, \exp(-\nu/\nu_{\rm cut, C});\,\,  \nu > \nu_{\rm t}
\end{equation}
Fig. \ref{sketch} illustrates graphically the chosen interpolating model, together with
the parameters.
The data used to find the best interpolation to the data are all the data present in 
the archive of ASDC, plus the averaged data (for each $\gamma$--ray luminosity bin)
found directly from the fluxes listed in the 3LAC catalog.
We believe that this description, while sufficiently accurate, is simpler than
the corresponding phenomenological description given in F98.
The constants $A$, $B$, $C$ are obtained requiring:

\begin{figure*} 
\vskip -0.7 cm
\hskip -1. cm
\includegraphics[height=24cm]{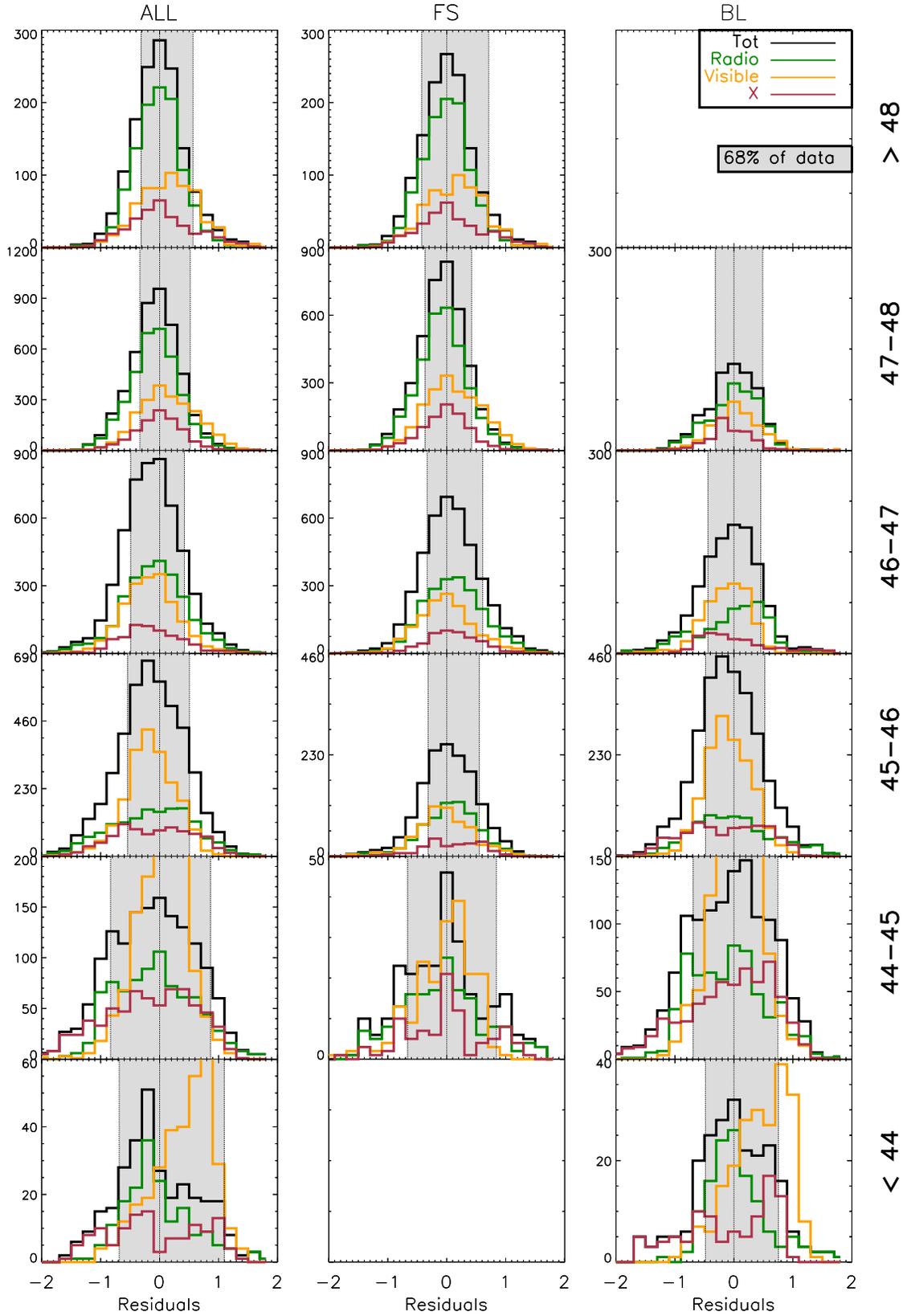}
\vskip -1.2 cm
\caption{Residuals (data--model) distributions for all blazars
and for FSRQs and BL Lacs only.
We show separately the residuals in the radio (green line), visible (orange) 
and X--ray (red) data, and their sum (black).
For the luminosity bins [$\log L_\gamma>48$] and  [$47<\log L_\gamma<48$] 
the accretion disc contributes to the optical luminosity for FSRQa.
Therefore we do not consider the optical luminosity distribution for
the total dispersion.
The same occurs for the two lowest luminosity bins of all blazars
due to the emission of the host galaxy.
} 
\label{isto}
\end{figure*}

%
\begin{itemize}
\item The radio spectrum and $L_{\rm S+C}(\nu)$ connect at $\nu_{\rm t}$;
\item at $\nu_{\rm S}$ (the peak of the synchrotron spectrum), the
luminosity is $L_{\rm S}(\nu_{\rm S})$;

\item at $\nu_{\rm C}$ (the peak of the inverse Compton spectrum), the
luminosity is $L_{\rm C}(\nu_{\rm C})$.
This is parametrized giving the Compton dominance parameter CD,
namely the ratio of the $\nu L(\nu)$ Compton and synchrotron luminosities.
\end{itemize}

In total, we have 11 parameters, all univocally related to observables.
They are:
\begin{itemize}
\item the three typical frequencies $\nu_{\rm t}$ (self--absorption), 
$\nu_{\rm S}$ (peak frequency of the synchrotron spectrum), 
$\nu_{\rm C}$ (peak frequency of the high energy spectrum);

\item the two cut--off frequencies $\nu_{\rm cut, S}$ and $\nu_{\rm cut, C}$
of the synchrotron and high energy spectra, respectively;

\item the four slopes $\alpha_{\rm R}$ (radio), $\alpha_1$ (connecting $\nu_{\rm t}$ with $\nu_{\rm S}$),
$\alpha_2$ (the slope after the synchrotron and after the high energy peak), $\alpha_3$ 
(the slope before the high energy peak).
When the scattering occurs in the Klein--Nishina regime, the slope of the synchrotron
after the peak is not equal to the slope above the Compton peak. 
This is a key point for BL Lacs detected at TeV energies, but is less important
in the 0.1--100 GeV band for BL Lacs detected by {\it Fermi}. 
Therefore for simplicity, we assume that the two slopes are equal. 

\item the two $\nu L(\nu)$ luminosities at the synchrotron and the high energy peaks.

\end{itemize}

We found that the same $\alpha_{\rm R}=-0.1$ can describe all sources, so the free parameters 
become 10.
Furthermore, the high energy cut--offs are not well determined, but are not 
very important for the description of the SED. 
Consider also that all parameters are observable quantities.

\begin{figure*} 
\vskip -1.4 cm
\includegraphics[height=22.1cm]{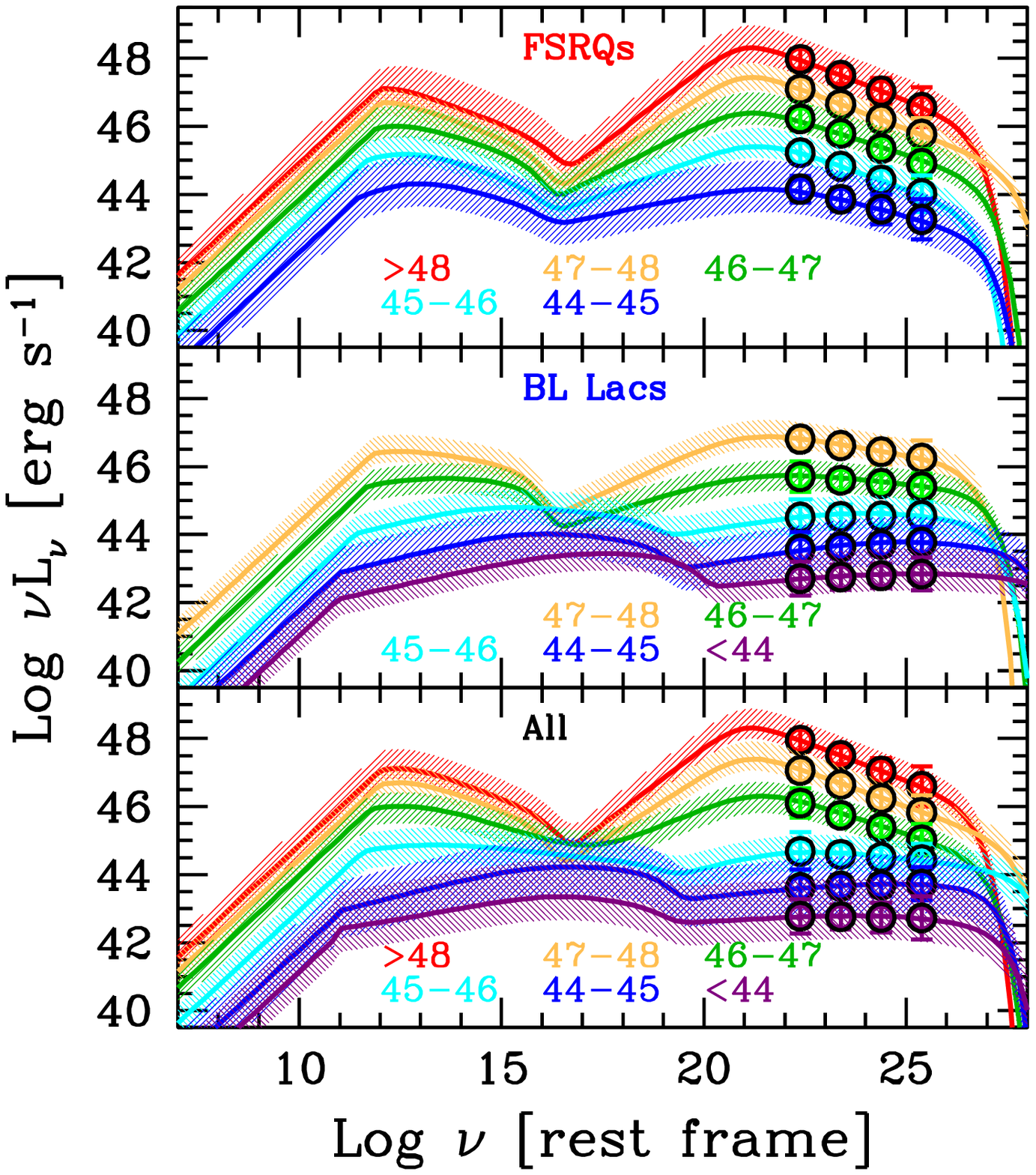}
\vskip -1.4  cm
 \caption{
The new phenomenological sequence, obtained for different bins of the logarithm of 
the $\gamma$--ray luminosities (indicated by the labels)
for FSRQs only (top), BL Lacs (mid) and for all sources (bottom), 
according to the parameters listed in Table \ref{para}.
For FSRQs, the spectral shape changes only slightly as the bolometric luminosity
changes, except for the Compton dominance. 
Therefore the most remarkable change is in the X--ray slope, becoming harder 
as the luminosity increases.
The synchrotron peak frequency $\nu_{\rm S}$ often coincides with the self--absorption 
frequency, possibly hiding a change in $\nu_{\rm S}$.
The Compton peak frequency $\nu_{\rm C}$ changes only weakly, if at all.
For BL Lacs both the synchrotron and the Compton peak frequencies become  smaller
(and the Compton dominance as increases) increasing the bolometric luminosities, 
as in the original blazar sequence.
The $\gamma$--ray slope is harder for less luminous BL Lacs, but not as much as in the 
original sequence (we remind that F98 had only three EGRET--detected BL Lacs in the 
less luminous bin).
Merging together FSRQs and BL Lacs we do see a clearer sequence, as in F98.
} 
\label{seq2}
 \end{figure*}

\subsection{Finding the best interpolation}

From the ASDC archive, we extract the data points of each source.
The Catalina fluxes and the USNO data, often present,
show fluxes outside the range of the other data at the same frequency.
Given their large uncertainty, we disregard them.
The archival data are of different quality: there are sources
observed multiple times and with an excellent frequency coverage, while others 
have a much worse data coverage.
Therefore, before considering all data, we averaged the fluxes
at the same frequency for the same source, not to over--weight these
more observed sources when finding the best fit.
For the $\gamma$--ray data the 3LAC catalog gives directly the flux and
spectral index for each source averaged over 4 years of observations.
We compute the average $\gamma$--ray data  for all blazars in the same $\gamma$--ray luminosity bin, 
and the  dispersion around the mean of these data. 

In each $\gamma$--ray luminosity bin  the best interpolation is found 
considering the scatter of the data points, in the following way.
We construct a model, according to Eq. \ref{eq1} and Eq. \ref{eq2}, 
and compute the residuals (data--model) of all points, divided  
into three wide spectral bands: the radio (up to $10^{12}$ Hz),
the IR--optical--UV (between $10^{12}$ and $10^{16}$ Hz) and the X--ray 
(between $10^{16}$ and $10^{20}$ Hz), as shown in Fig. \ref{sketch}.
There are specific luminosity bins where the accretion disc (for FSRQs at high luminosities)
or the host galaxy (at low luminosities) contribute to the optical flux.
In this case we exclude the residuals in the optical range in the computation 
of the best interpolation.
The best interpolation is found when the median of all residuals is zero.

\begin{figure} 
\vskip -0.7  cm
\hskip -1.8 cm
\includegraphics[height=12.3cm]{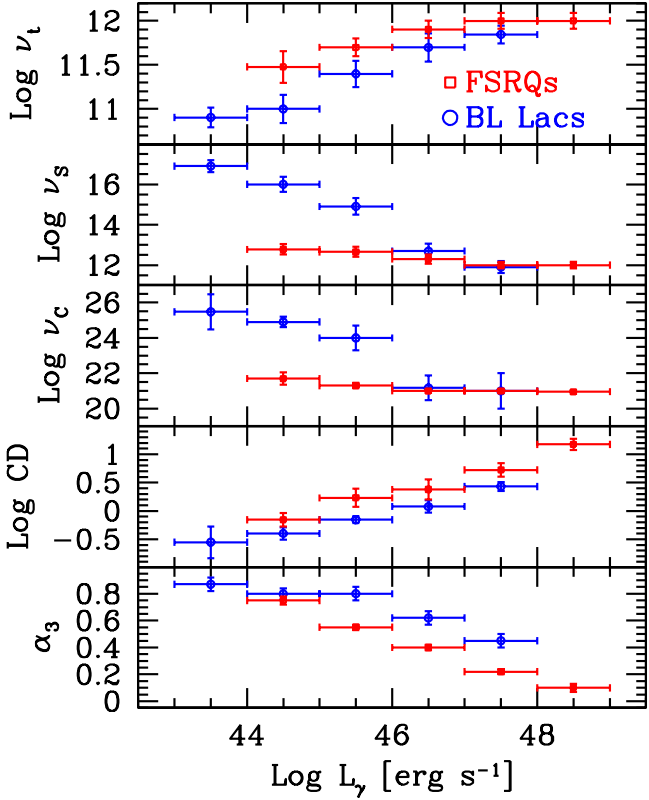}
\vskip -0.5 cm
\caption{From top to bottom:
synchrotron self--absorption frequency ($\nu_{\rm t}$),
synchrotron peak frequency ($\nu_{\rm S}$),
high energy peak frequency ($\nu_{\rm C}$),
Compton Dominance (CD) and value of $\alpha_3$
as a function of the $\gamma$--ray luminosity in the LAT band.
The FSRQs have constant peak frequencies, but become more
Compton dominated and have a harder $\alpha_3$--slope (for them it
coincides with the slope in the X--ray band) as the $\gamma$--ray 
luminosity increases.
BL Lacs, in addition, also becomes ``redder" (smaller peak frequencies) as the
$\gamma$--ray luminosity increases.
} 
 \label{vp}
 \end{figure}
\begin{figure} 
\vskip -0.7  cm
\hskip -1.8 cm
\includegraphics[height=12.3cm]{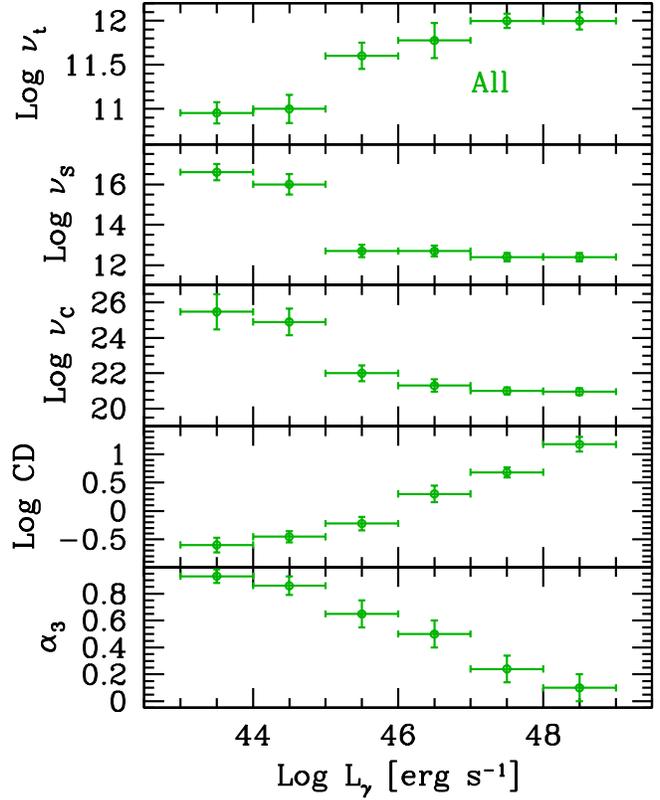}
\vskip -0.5 cm
\caption{
Same as Fig. \ref{vp}, but for all blazars together. 
Note that in this case the trends are even more pronounced
than the sequence for BL Lacs only.  
} 
\label{vpall}
\end{figure}

In Fig. \ref{isto} we show the histograms of the residuals for all luminosity bins
for FSRQs, for BL Lacs and for all blazars together.
The grey shaded areas and the vertical lines in these histograms show the ranges containing
the 68\% of the sources. 
There are two ways to calculate it. 
The first is to start from the median, and calculate the symmetric
(equal to the left and the right of the median) range of residuals containing
the 68\% of the sources.
This is appropriate when the distribution are symmetric around the median.
In our case, the distributions are often asymmetric and we prefer to
calculate where there are 16\% of the sources starting from the most negative value
of the residuals and the 16\% of the sources starting from the largest value.
This is why the values listed under the column ``$\sigma$"  in Table \ref{para}
are not symmetric.
The width of the stripes shown in Fig. \ref{seq2} and Fig. \ref{newold}
corresponds to these values.

All parameters are listed in Table \ref{para}: separately for FSRQs and BL Lacs
and also for all blazars together.
The corresponding SEDs are shown in Fig. \ref{seq2}.
In this figure we also show the K--corrected average luminosities
in the $\gamma$--ray (from 0.1 to 100 GeV).
The errobars of these points corresponds to the dispersion around
the mean (at 1 $\sigma$).

\section{Results}

Fig. \ref{seq2} is the main result of our study.
It shows the interpolating models for FSRQs, for BL Lacs and
for  FSRQS+BL Lacs (from top to bottom).
Fig. \ref{vp} and Fig. \ref{vpall} show how the
most important parameters change with the $\gamma$--ray luminosity.
The errors shown in Fig. \ref{vp} and Fig. \ref{vpall} have been derived
in the following way.
First we calculated the error $\Delta M$ of the median of the residuals of the best fit,
which we approximate as  $\Delta M= \sigma/\sqrt{N}$ (Taylor 1997),
where $\sigma$ is the dispersion of the residuals around the median, 
and $N$ is the number of data points.
Then we change one parameter of interests, until the resulting
new median deviates by $\Delta M$.
Some parameters, like $\nu_{\rm t}$  or the Compton dominance,
do not affect  the entire frequency range, but only a specific band.
We then consider only that band when calculating $\Delta M$.

\begin{figure*}
\hskip 1 cm 
\vskip -3.4 cm
\includegraphics[height=16cm]{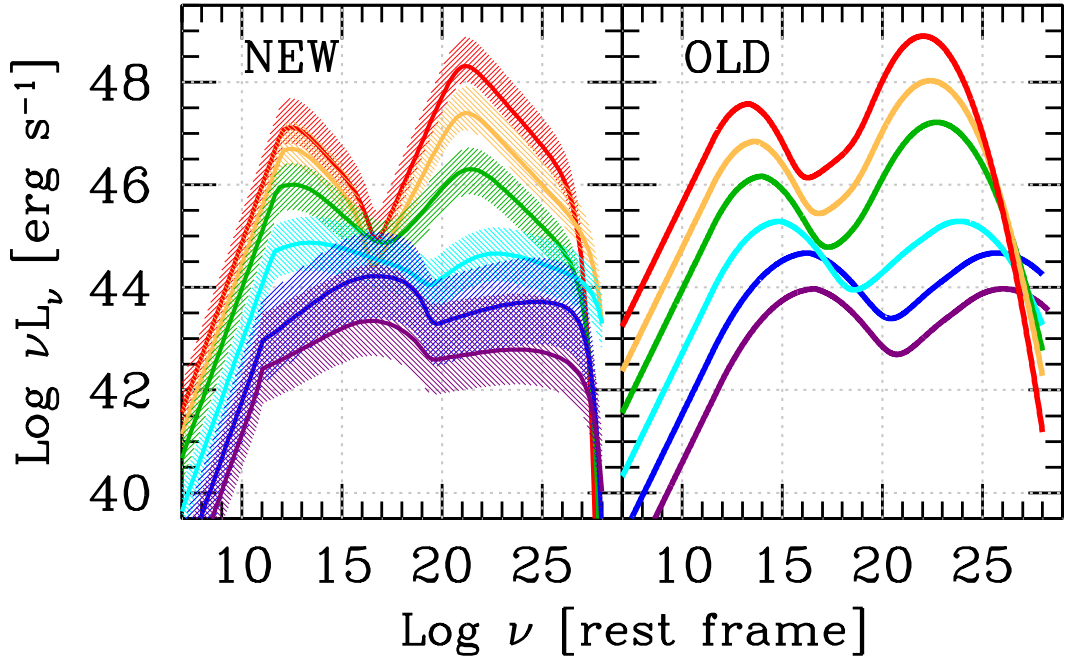}
\vskip -3.8 cm
 \caption{
Comparison between the new and the original blazar sequence for all blazars.
Note that the original blazar sequence considered 5 radio luminosity bins, while
the new one considers bins in the $\gamma$--ray band.
} 
\label{newold}
 \end{figure*}

\vskip 0.2 cm
\noindent
{\it FSRQ sequence ---}
Focusing first on FSRQs, we note that the overall shape of the SEDs
does not change much, except for the Compton dominance, that
increases with the $\gamma$--ray (hence bolometric) luminosity,
as in the original blazar sequence (see Fig. \ref{seq1}).
Contrary to the overall sequence, the $\gamma$--ray slope is almost 
constant, as well as the peak frequencies of the synchrotron and the 
high energy components (see Table \ref{para}).
The synchrotron peak frequency $\nu_{\rm S}$ is of the same order 
of the self--absorption frequency $\nu_{\rm t}$ and this
inhibits somewhat our ability to find a trend (i.e. if $\nu_{\rm S}$
becomes smaller than $\nu_{\rm t}$ we may not clearly notice it).
The ``Compton" peak frequency $\nu_{\rm C}$ varies slightly from $5\times 10^{21}$ to 
$9\times 10^{20}$  Hz (i.e. from $\sim$21 to $\sim$4 MeV) for an increase in luminosity  
of 4 orders of magnitude.
Albeit small, this change can have important implications for the planning of future 
missions such as e--Astrogam (e.g. De Angelis et al. 2016) aimed to explore 
the energy range around 1 MeV. 
There are some indications, in fact, that {\it Fermi} misses to detect the most powerful
blazars, that are instead detected by {\it Swift}/BAT (see Ghisellini et al. 2010).
Since the most powerful blazars are at high redshifts, 
this can be due to the K--correction, shifting the observed high energy peak
to small energies, such that the steep high energy tail beyond the peak
becomes not detectable by {\it Fermi}/LAT.
But another, additional, reason is that also the rest frame peak frequency 
could be indeed smaller when increasing the total luminosity.

The most notable trends for FSRQs are the sequence in Compton dominance
smoothly increasing with luminosity (from 0.5 to 15), and the slope in
X--rays (described for FSRQs by $\alpha_3$), becoming harder with luminosity
(from $\alpha_3=0.75$ to 0.1).
This is easily explained: since the $\gamma$--ray luminosity increases 
more than the synchrotron one, the slope between the end of the synchrotron
and the ``Compton" peak must necessarily become harder.
The above properties can find an easy explanation in terms of a nearly constant
radiative cooling rate, as discussed in the next section.

\vskip 0.2 cm
\noindent
{\it BL Lac sequence ---} 
For BL Lacs there is a remarkable trend for the entire SED, that changes
by changing the bolometric luminosity.
The peak frequencies become smaller as the luminosity increases:
$\nu_{\rm S}$ goes from $\sim 10^{17}$ Hz to $\sim 10^{12}$ Hz
increasing $L_\gamma$ by 4 orders of magnitude.
Similarly, $\nu_{\rm C}$ decreases by more than 4 orders of magnitude,
with most of the change occurring around $L_\gamma\sim 10^{46}$ erg s$^{-1}$.
This is due to the change of the $\gamma$--ray spectral index $\alpha_\gamma$, very close to unity.
Although $\alpha_\gamma$ changes slightly, when it becomes a little harder than
unity it changes $\nu_{\rm C}$ by a large amount.
The Compton dominance changes by one order of magnitude in the entire luminosity range
(from $\sim$0.3 to $\sim$3), indicating that in BL Lacs the synchrotron and the high energy
components are almost equal.
There is a clear trend in the spectral index $\alpha_3$, as shown in Fig. \ref{vp}.
Due to the large shift of $\nu_{\rm S}$, this index is the X--ray spectral index 
at high luminosities, becoming the $\gamma$--ray spectral index at low luminosities.
Overall, the trends are less pronounced than in the original blazar sequence,
especially for the $\alpha_\gamma$ index, never so hard as in the F98 paper.

\vskip 0.2 cm
\noindent
{\it Blazar sequence ---}
When we put together all blazars, the sequence becomes more evident, and more
similar to the original one.
This is because, at high luminosities, the properties of the average SED are dominated
by FSRQs, while BL Lacs dominate the average SED at low $L_\gamma$.
To illustrate this point, Fig. \ref{newold} compares the ``new" and the ``old" sequences. 
Bear in mind that the original sequence was constructed considering bins
of radio luminosities, while the new one divides blazars according to the
$\gamma$--ray luminosity.
The main difference concerns the Compton dominance at low luminosities, now smaller,
and the $\gamma$--ray slope, that now is barely harder than unity at low luminosities. 
Another difference concerns the synchrotron peak, changing smoothly in 
the old sequence, and more abruptly in the new one, around $L_\gamma\sim 10^{45}$
erg s$^{-1}$, as shown in Fig. \ref{vpall}.

\vskip 0.2 cm
\noindent
{\it Different black hole masses confuse the sequence ---}
As discussed in \S 2, the original sequence very likely
concerned blazars with large black hole masses
(of both the BL Lac and FSRQ families).
Now, that we have an improved sensitivity at all frequencies,
we start to detect blazars with black holes of smaller masses.
This is particularly so at low luminosities, where we now
find intermediate power FSRQs (at $z\sim 0.2$), detected at fluxes that could
not be reached before, and relatively nearby BL Lacs (with $z<0.1$)
of intermediate power as well, but that can have 
large black hole masses.
In the $\gamma$--ray band, they can have the same luminosity, and therefore they
are grouped in the same luminosity bin.
We suggest that this is the main reason of the different SED of 
FSRQs and BL Lacs below $L_\gamma\sim 10^{45}$--$10^{46}$ erg s$^{-1}$.

\begin{figure} 
\vskip -0.7 cm
\hskip -0.5 cm
\includegraphics[height=9.5cm]{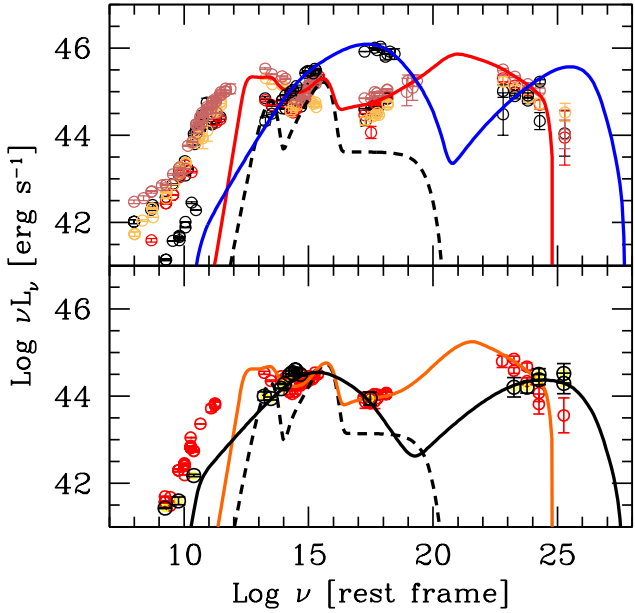}
\vskip -0.5 cm
\caption{
{\it Top panel}: 
The SED of a few FSRQs  (PKS 1352--104, $z$= 0.33; PKS 1346--112, $z$= 0.34; 
S4 0110+49, $z$= 0.389 and  5BZQ 1153+4931, $z$= 0.334) 
compared with the SED of 1ES 0502+675 ($z$= 0.34), a blue BL Lac object.
All blazars have the same $L_\gamma$ (they belong to the same bin
$45 < \log (L_\gamma/\rm erg \, s^{-1} )<46$), but they very different SEDs. 
The red (for the FSRQs) and blue (for the BL Lac object) 
lines correspond to a model (Ghisellini \& Tavecchio 2009)
that tries to explain the IR to GeV emission, including the thermal components.
The black hole mass found for the FSRQs is around $10^8 M_\odot$ and 
the disc luminosity is $L_{\rm d}=0.1 L_{\rm Edd}$.
For BL Lac we assumed a black hole mass $10^9 M_\odot$ and $L_{\rm d}=10^4 L_{\rm Edd}$.
{\it Bottom panel:} the same occurs also at smaller luminosities.
In this figure, 
PMN 0017--0512 ($z$= 0.227, orange line)
is an FSRQs with a visible accretion disc component, while 
PMN 2014--0047  ($z$= 0.23, black solid line) is a BL Lac object.
FSRQs with black hole of low mass and disc luminosity
$L_{\rm d}\sim 0.1 L_{\rm Edd}$ 
have jets of relatively small luminosities, but a SED looking 
alike the most powerful blazars.
On the contrary, BL Lacs with a large black hole mass but
with $L_{\rm }\ll 10^{-2} L_{\rm Edd}$ may have 
the same $L_{\gamma}$ but a completely different SEDs.
}
\label{lowmass}
\end{figure}

This is illustrated by the examples shown in Fig. \ref{lowmass}.
Let us discuss the top panel first, comparing the 
SEDs of a few FSRQs of similar spectral shape and equal $L_\gamma$
(PKS 1352--104, $z$= 0.33; 
PKS 1346--112, $z$= 0.34; 
S4 0110+49, $z$= 0.389;  
5BZQ 1153+4931, $z$= 0.334)
with the SED of a BL Lac in the same $L_\gamma$ bin (1ES 0502+675, $z$= 0.34, black points).
Fig. \ref{lowmass} shows also two models (solid red line for the FSRQs and blue line for the BL Lac) 
that can interpret the sub--mm to GeV emission, including the thermal part.
This model is described in Ghisellini \& Tavecchio (2009).
The upturn in the optical of the FSRQs (red line) can be interpreted
as due to the contribution of the accretion disc.
Together with the broad line luminosity, we can estimate the black hole mass ($10^8 M_\odot$) 
and the accretion rate (0.1 $L_{\rm Edd}$).
For the BL Lac we do not have any thermal emission to derive the 
black hole mass and disc luminosity, but the shown fit reports the
case of a black hole mass of $10^9 M_\odot$ with a disc emitting at $10^{-4}$ 
of the Eddington limit.
The disc of the BL Lac is radiatively inefficient 
(see e.g. Narayan, Garcia \& McClintock, 1997)
and cannot photo--ionize the BLR to sustain broad lines of
luminosities comparable to the ones of a standard quasar.
The lack of external photons to be scattered at high frequencies
implies less $\gamma$--ray emission (i.e. less Compton dominance),
less radiative cooling and a larger average electron energy, resulting in a bluer spectrum
(large $\nu_{\rm S}$ and $\nu_{\rm C}$).
As a result, the overall SED of the two blazars is completely different,
including the $\alpha_\gamma$ slope, and yet the average $L_\gamma$ is the same.
This can also be seen by comparing, in Fig. \ref{seq2}, the overall phenomenological SED
of BL Lacs and FSRQs at intermediate/small luminosities.

The bottom panel shows the comparison of two less powerful objects: 
one is a FSRQ (PMN 0017--0512, $z=$0.227, red points) and the other is a BL Lac 
(PMN 2014--0047, $z$=0.23, black points).
The two sources have comparable radio fluxes. 
Again, the FSRQ shows the upturn in the optical produced by the disc.
This, together with the broad line luminosity, can fix 
the black hole mass ($4\times 10^7 M_\odot$) and a disc luminosity of $0.1 L_{\rm Edd}$.
For the BL Lac we again use a black hole mass of $10^9 M_\odot$ with a disc emitting at
the $10^{-4}$ of the Eddington limit.

These cases suggest that when different black hole masses are present,
the blazar sequence is ``polluted" by objects of small black hole mass,
relatively small luminosity and a ``red" look, that fall in the same luminosity
bin of  ``blue" BL Lacs. 
On one hand, this implies that the blazar sequence is not controlled by the
observed luminosity only, but rather by the Eddington ratio.
On the other hand, by dividing the objects on the basis of their
luminosity, as done here (and also in the original F98 sequence)
and considering all blazar together, this implies 
a large dispersion of points around the median spectrum, especially
so for the medium and low luminosity bins, as reported in Table \ref{para} and as 
can be seen in Fig. \ref{seq2} and in Fig. \ref{newold}.

\vskip 0.2 cm
\noindent
{\it Disc emission ---}
A shown in Fig. \ref{bbb}, there is a sizeable fraction of FSRQs 
showing a clear presence of thermal disc emission in the optical.
This is flagged by an upturn of the total spectrum, since
the synchrotron peak frequency is, on average, in the mm band
(see Table \ref{para}: $\langle \nu_{\rm S}\rangle \sim 2.5\times 10^{12}$ Hz 
for FSRQs of all luminosities).
The fraction of disc whose emission is clearly visible increases
with $L_\gamma$.
This can be due to fact that the disc component
becomes more visible at large redshifts (hence at large luminosities),
since in this case the peak of the disc emission (in the UV, rest frame) 
falls into the observed optical band.

Another interesting point concerns the relation between the disc and the
jet luminosity. 
This point has been discussed in Ghisellini (2016): here we briefly
mention that when we normalize the entire SED to the peak of the 
disc emission, we obtain a reduced scatter in the other bands,
especially in the radio and in the X--rays
(in the $\gamma$--rays the variability amplitude is so large to hide
the reduction of the scatter).
This can be taken as a model independent evidence that 
the disc and the jet luminosity are related.

\section{Discussion}

The following are the main findings of our study:
\begin{itemize}
\item The blazars detected by {\it Fermi}/LAT of known redshift
do form a sequence, with the same general properties of the original blazar 
sequence found by F98.
\item The differences between the new and the old sequence
are explained by the fact that the objects detected by {\it Fermi} 
do not represent ``the tip of the iceberg" in terms of their 
$\gamma$--ray luminosity, as was the case for EGRET and
for the original blazar sequence.
\item When considering BL Lacs and FSRQs separately, we discover
that FSRQs do form a sequence, but only in Compton dominance
and in the X--ray slope. 
They do not become redder when more luminous, while BL Lacs do.
\item At high luminosities and redshift, the accretion disc
become visible in FSRQs.
\item We have considered only the blazars detected by {\it Fermi}/LAT.
There are other blazars, at both extremes of the luminosity range,
that are not detected by {\it Fermi}.
Most notably, extreme BL Lacs, with their high energy emission
peaking in the TeV band (e.g. Bonnoli et al. 2015), 
and very powerful FSRQs, whose high energy
peak lies in the MeV band (Ghisellini et al. 2010).
\end{itemize}

\subsection{A physical insight}

The fact that FSRQs follow a clear trend in Compton dominance,
while the peak frequencies $\nu_{\rm S}$ and $\nu_{\rm C}$
are almost constant calls for an explanation.
We suggest that an important ingredient to interpret this behavior
is given by the radiative cooling rate of FSRQs.
A key assumption is that most of the cooling occurs within the
BLR and/or the molecular torus, and that the inverse Compton (IC)
scatterings off this external radiation is dominant, or at 
least competitive with respect to the synchrotron cooling.
The IC cooling rate is proportional to the radiation energy density 
$U^\prime_{\rm ext}$ as measured in the comoving frame of the emitting jet region,
moving with a bulk Lorentz factor $\Gamma$.
This is of the order:
\begin{equation}
U^\prime_{\rm ext }\, \sim \, \Gamma^2 \, \left[ U_{\rm BLR} + U_{\rm T} \right]
\end{equation}
where the unprimed quantities are measured in the black hole frame.
$U_{\rm BLR}$ and $U_{\rm T}$ are the radiative energy densities
due to the BLR and the torus, respectively.

If the radius of the broad line region $R_{\rm BLR}$ scales as  
$R_{\rm BLR} \sim 10^{17} L_{\rm d, 45}^{1/2}$ cm, 
we have  (see e.g. Ghisellini \& Tavecchio 2009):
\begin{equation}
U_{\rm BLR }\, = \, { a L_{\rm d} \over 4\pi R^2_{\rm BLR} c} 
\, \sim {1\over 12\pi} \qquad {\rm erg \, cm^{-3}}
\end{equation}
where we have assumed a covering factor $a=0.1$ for the broad line clouds.

A similar argument can be done for the relevant distance of the molecular torus.
We can assume that the absorbing dust survives at a temperature $T_{\rm T}<10^3$ K.
We can also assume that the torus re-emits all the disc radiation it intercepts.
This is a fraction $f$ of $L_{\rm d}$, that depends on the geometry of the torus itself.
For simplicity, and very crudely,  let us assume that its shape is similar
to a portion of a sphere of radius $R_{\rm T}$ that is also the distance from the
black hole. 
Assuming that $f 4\pi R^2_{\rm T}$ is the total surface of the torus, we have:
\begin{equation}
4\pi f R^2_{\rm T}\sigma_{\rm SB} T^4_{\rm T}   =  f L_{\rm d}
\, \to \,
R_{\rm T}   \sim 1.2 \times 10^{18}     L^{1/2}_{\rm d, 45}  T_{\rm T, 3}^{-2} 
\,\, {\rm cm} 
\end{equation}
Since  $R_{\rm T}\propto L^{1/2}_{\rm d}$, also the energy density $U_{\rm T}$ 
produced by the torus is constant, as long as we are at a distance smaller than 
$R_{\rm T}$ from the black hole:
\begin{equation}
U_{\rm T }\, \sim  \,  {0.07 f \over 12 \pi}
\, \qquad {\rm erg \, cm^{-3}}
\end{equation}
We can conclude that in the comoving frame, $U^\prime_{\rm ext}\propto \Gamma^2$,
therefore it is nearly constant if the $\Gamma$--factor is approximately
the same in different sources.
In this case the cooling rate is the same in FSRQs of different power.

For BL Lacs, instead, the main radiation mechanism for the high energy hump
is the synchrotron Self--Compton process. 
It strongly depends upon the synchrotron radiation energy density in the 
comoving frame that is an increasing function of the observed luminosity.
Therefore the cooling is not constant, but it is more severe in more
powerful BL Lacs.
As a consequence, the energy of the electrons emitting at the SED peak
could be smaller (due to the larger cooling) in high luminosity BL Lacs,
and both $\nu_{\rm S}$ and $\nu_{\rm C}$ would be smaller.
In FSRQs, instead, the same cooling rate implies the same relevant electron energies 
and constant peak frequencies.

For FSRQs we have to explain also why the Compton dominance increases 
with total power.
Since $U^\prime_{\rm ext}$ is roughly constant (for constant $\Gamma$),
the only possibility is that the magnetic energy density 
$U^\prime_{\rm B}$ {\it decreases} with luminosity.
The ratio $U^\prime_{\rm ext}/U^\prime_{\rm B}$, in fact, determines the
value of the Compton dominance.
Since 
$\nu_{\rm S} \propto B^\prime$, we expect that more powerful
FSRQs have a smaller $\nu_{\rm S}$.
At the same time, the self--absorption frequency $\nu_{\rm t}$ 
is increasing with synchrotron luminosity:
\begin{equation} 
\nu_{\rm t}\, \propto \, B\left[K {R\over B}\right]^{2/(p+4)}  
\end{equation}
(e.g. Rybicki \& Lightman 1979),
where $R$ is the size of the innermost emitting region, $K$ is the density of 
the relativistic  emitting electrons and $p$ is the slope of the electron energy 
distribution $N(\gamma)=K\gamma^{-p}$.
Assuming the same $R$, the synchrotron luminosity $L_{\rm S}\propto B^2 K$,
so that 
\begin{equation} 
\nu_{\rm t}\, \propto\, B \left[ {L_{\rm S} \over B^3}\right]^{2/(p+4)} 
\end{equation}
For $p=2$ we obtain $\nu_{\rm t} \propto L_{\rm S}^{1/3}$.
Therefore, at the high luminosity extreme, the flux emitted at $\nu_{\rm S}$
could be self--absorbed, and what we observe as the synchrotron peak
frequency is instead $\nu_{\rm t}$.
This is what we find, as can be seen in Table \ref{para} and
Fig. \ref{vp}.
Note, in fact, that FSRQs in the highest two luminosity bins
have $\nu_{\rm t}\sim \nu_{\rm S}$.

Since  both the relevant electron energy and the 
seed photon frequency are approximately constant with luminosity, 
also the peak of the high energy component $\nu_{\rm C}$ is expected
to be approximately constant, as observed and shown in Fig. \ref{vp}.

\begin{figure} 
\vskip -0.5 cm
\includegraphics[height=9cm]{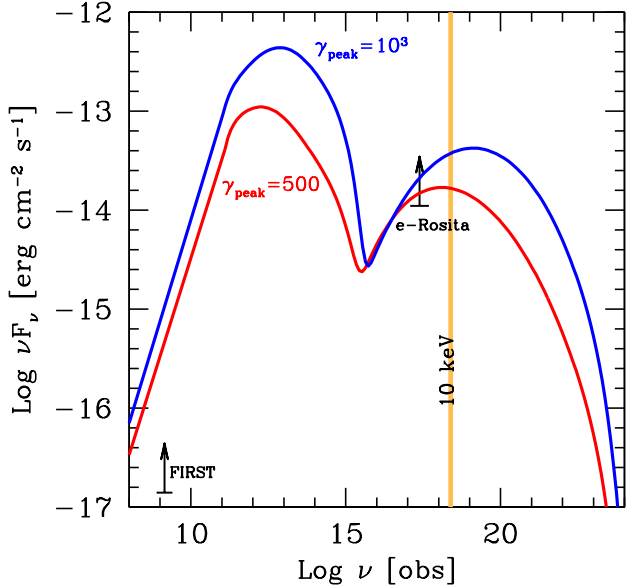}
\vskip -0.5 cm
\caption{Example of SED derived according to the prescriptions
of Giommi et al. (2012).
We in fact assume that the Doppler factor $\delta=15$, 
the magnetic field $B=0.15$ G.
For the blue line we assume $\gamma_{\rm peak}=10^3$ and $z$=0.2,
for the red line $\gamma_{\rm peak}=500$ and $z$=0.25.
The two black arrows refer to the limiting sensitivity
of the FIRST survey (i.e. 1 mJy at 1.4 GHz) and
the foreseen sensitivity of the all sky suvey of {\it eROSITA}
(i.e. $1.1\times 10^{-14}$ erg cm$^{-2}$ s$^{-1}$ in the 0.5--2 keV band). 
}
\label{giommi}
\end{figure}

\subsection{Blazar sequence vs the ``simplified scenario"}
\label{simplified}

As discussed in the introduction, Giommi et al. (2012) proposed the ``simplified scenario" 
for blazars, completely alternative to the blazar sequence.
According to these authors, the blazar sequence is indeed what we observe,
but only as a result of selection effects.
In reality, there is no physical link between the luminosity and the overall shape of a blazar.
Both at high and low luminosities we have both red and blue blazars.
In this scenario
the distribution of $\gamma_{\rm peak}$ is assumed to peak at $\gamma_{\rm peak}\sim 10^3$,
and it is asymmetric, with a high energy tail broader than the low--energy one
(see Fig. 4 of Giommi et al. 2012)
Furthermore the radio luminosity function is $\Phi (L) \propto L^{-3}$,
the magnetic field is assumed to be $B=0.15$ G for all sources, and the 
mean Doppler factor distribution has a mean value of $\langle\delta\rangle=15\pm 2$. 
The immediate consequence is that there exist many BL Lacs of low power with a red SED. 
These are the ones with $\gamma_{\rm peak} \sim 10^3$ or smaller.
We do not see them yet because, being at low luminosities, we do not have
survey deep enough to let them emerge.
These objects should have a high energy, synchrotron self Compton
spectrum peaking in the 1--10 keV band.
Fig. \ref{giommi} shows two examples of the expected spectrum,
with $\gamma_{\rm peak}\sim 10^3$ (blue line) and $\gamma_{\rm peak}=500$ (red line).
We have $\nu_{\rm S} \sim 8\times 10^{12} \gamma^2_{\rm peak, \, 3}$ Hz
and $\nu_{\rm C} \sim  \gamma^2_{\rm peak} \nu_{\rm S}\sim 8\times 10^{18}\gamma^4_{\rm peak, \, 3}$ 
Hz ($\sim$ 33 keV).
According to these examples, there should be BL Lacs already detected in the 
radio, at moderate (but still unknown) redshifts, with no X--ray detection yet.
On the other hand, with the advent of {\it eROSITA} (Merloni et al. 2012), these objects would be 
detected in X--rays, helping to probe their BL Lac nature.

\subsection{Bias against BL Lacs?}

Given the difficulties to measure the redshift of BL Lacs, their 
redshift distribution probably does not reflect the true one, 
and it is likely biased against high redshift BL Lacs. 
Rau et al. (2012), observed a sample of 106 {\it Fermi}/LAT blazars  with UVOT (onboard {\it Swift}) and
the Gamma-Ray Optical/Near--infrared Detector (GROND; Greiner et al. 2008).
This allowed them to derive an upper limit to the redshifts of 66 sources, and
a photometric redshift estimate (through the drop--out technique) for 8 blazars, 6 of which
were found at $z>1.3$.
However, the absence of emission lines in these blazars (leading to their classification
as BL Lac objects) was discussed by Padovani, Giommi \& Rau (2012), who studied
the 4 sources having a 	``blue" optical spectrum, concluding that 
these sources were FSRQs whose optical emission was so enhanced to swamp their emission lines 
(see also Ghisellini et al. 2012, who confirmed this conclusion).

More recently, Mao \& Urry (2016) and Mao (2017) considered a large sample of BL Lacs without
redshift and compared their properties against a very large sample of BL Lacs
of known redshift.
They find that ``blue" BL Lacs at high redshift (therefore very powerful), if
they exist, are very rare.

\subsection{Blazar sequence and $\gamma$--ray background}

In a recent paper, Giommi \& Padovani (2015) computed 
the contribution of blazars to the $\gamma$--ray background
according to their ``simplified scenario" and also according to the
old F98 blazar sequence.
For the latter,
they used their radio luminosity function $\Phi(L) \propto L^{-3}$,
and for any radio luminosity they assigned the corresponding $\gamma$--ray one
as indicated in the F98 paper (specifically, their Fig. 12).
They found that the blazar contribution to the $\gamma$--ray background, calculated
in this way, severely exceeds the observed $\gamma$--ray background,
while using the prescription of the ``simplified scenario" they obtained
a good agreement.
However, as discussed above, the old F98 blazar sequence described the ``tip of the iceberg"
of the $\gamma$--ray emission of blazars, that largely exceeded the average one.
We have re--computed the $\gamma$--ray background according to the new sequence,
and found a good agreement (Bonnoli et al. in prep.).

\section*{Acknowledgements}
We thank the referee for his/her constructive comments.
Part of this work is based on archival data, software or online services provided by the 
ASI SCIENCE DATA CENTER (ASDC).  
We acknowledge financial contribution from the agreement ASI--INAF I/037/12/0
(NARO 15), from the CaRiPLo Foundation and the
regional Government of Lombardia for the project ID 2014-1980 ``Science and technology
at the frontiers of $\gamma$--ray astronomy with imaging atmospheric Cherenkov Telescopes",
and from PRIN INAF 2014.


\begin{thebibliography}{99}

\bibitem[]{} Ackermann M., Ajello M., Atwood W.B. et al., 2015, ApJ, 810, 14 
           
\bibitem[]{} Ant\'on S. \& Browne I.W.A., 2005, MNRAS, 356, 225


\bibitem[]{} Bonnoli G., Tavecchio F., Ghisellini G. \& Sbarrato, T., 2015, MNRAS, 451, 611  

\bibitem[]{} Caccianiga A. \& Marcha M.J.M., 2004, MNRAS, 348, 937

\bibitem[]{} Calderone G., Ghisellini G., Colpi M. \& Dotti M., 2013, MNRAS, 431, 210 


\bibitem[]{} Chiaberge M. Capetti A. \& Celotti A., 1999, A\&A, 349, 77

\bibitem[]{} De Angelis A., Tatischeff V., Tavani M. et al., 2016, submitted to Experimental Astronomy
             (arXiv:1611.02232)  


\bibitem[]{} Donato D., Ghisellini G., Tagliaferri G. \& Fossati G., 2001, A\&A, 375, 739

\bibitem[]{} Elvis M., Plummer D., Schachter J., Fabbiano G., 1992, ApJS, 80, 257


\bibitem[]{} Fossati G., Maraschi L., Celotti A., Comastri A. \& Ghisellini G., 1998, 
             MNRAS, 299, 433 (F98)

	
\bibitem[]{} Ghisellini G., Celotti A., Fossati G., Maraschi L. \& Comastri A., 1998, MNRAS, 301, 451 

\bibitem[]{} Ghisellini G. \& Tavecchio F., 2008, MNRAS, 387, 1669  

\bibitem[]{} Ghisellini G. \& Tavecchio F., 2009, MNRAS, 397, 985    

\bibitem[]{} Ghisellini G, Maraschi L. \& Tavecchio F., 2009, MNRAS, 396, L105 
          
\bibitem[]{} Ghisellini G., Della Ceca R., Volonteri M. et al., 2010, MNRAS, 405, 387
           
\bibitem[]{} Ghisellini G., Tavecchio F., Foschini L. \& Ghirlanda G., 2011, MNRAS, 414, 2674 

             
\bibitem[]{} Ghisellini G., Tavecchio F., Foschini L., Sbarrato T., Ghirlanda G. \& Maraschi L.,
             2012, MNRAS, 425, 1371

\bibitem[]{} Ghisellini G., Tavecchio F., Maraschi L., Celotti A. \& Sbarrato T., 2014, Nature, 515, 376 

\bibitem[]{} Ghisellini G. \& Tavecchio F., 2015, MNRAS, 448, 1060


\bibitem[]{} Ghisellini G., 2016, Galaxies, 4, 36  (ArXiv:1609.08606)  

\bibitem[]{} Giommi P., Menna M.T. \& Padovani P., 1999, MNRAS, 310, 465 
            
\bibitem[]{} Giommi P., Piranomonte S., Perri M. \& Padovani P., 2005, A\&A, 434, 385

\bibitem[]{} Giommi P., Padovani P., Polenta G., Turriziani S., D'Elia V. \& Piranomonte S., 
             2012, MNRAS, 420, 2899 

\bibitem[]{} Giommi P., Padovani P., 2015, MNRAS, 450, 2404 

\bibitem[]{} Greiner J., Bornemann W., Clemens C. et al., 2008, PASP, 120, 405
             
\bibitem[]{} Jorstad S.G., Marscher A.P., Mattox J.R., Wehrle A.E., Bloom S.D., \& Yurchenko A.V.,
             2001ApJS, 134, 181 
            
\bibitem[]{} K\"uhr H., Witzel A., Pauliny--Toth I.I.K. \& Nauber U., 1981, A\&AS, 45, 367     

\bibitem[]{} Lister M.L., Aller M.F., Aller H.D. et al., 2013, AJ, 146, 120 

\bibitem[]{} Mao P., Urry C.M., Massaro F., Paggi A., Cauteruccio J., K\"unzel S.R., 2016, ApJS, 224, 26 
 
\bibitem[]{} Mao P. \& Urry C.M.,  2016, AAS 22724341 

\bibitem[]{} Mao P., 2017, PhD Thesis      
             
\bibitem[]{} Mattox J.R., Bertsch D.L., Chiang J. et al., 1996, ApJ, 461, 396 

\bibitem[]{} Merloni A., Predehl P., Becker W. et al., 2012,  {\it eROSITA science book}, arXiv:1209.3114, 2012. 

\bibitem[]{} Narayan R., Garcia M.R. \& McClintock J.E., 1997, ApJ, 478, L79
           
\bibitem[]{} Nieppola E., Tornikoski M. \& Valtaoja E., 2006, A\&A, 445, 441
 	
\bibitem[]{} Nieppola E., Valtaoja E., Tornikoski M., Hovatta T. \& Kotiranta M., 2008, A\&A, 488, 867     
    
\bibitem[]{} Padovani P. \& Urry C.M., 1992, ApJ, 387, 449 
               
\bibitem[]{} Padovani P., 2007, Ap\&SS, 309, 63 (astro--ph/0610545) 

\bibitem[]{} Padovani P., Giommi P. \& Rau A., 2012, MNRAS, 422, L48 

\bibitem[]{} Padovani P., Giommi P. \& Rau A., 2012, MNRAS, 422, L48

\bibitem[]{} Perlman E.S., Padovani P., Landt H., Stocke J.T., Costamante L., Rector T., Giommi P. 
             \& Schachter J.F.,  2001, ASPC, 227, 200 (astro--ph/0012185) 

\bibitem[]{} Raiteri C.M. \& Capetti A., 2016, A\&A, 587, A8 

\bibitem[]{} Rybicki G.B. Lightman A.P., 1979, Radative Processs in Astrophysics,
              Wiley \& Sons (New York)

\bibitem[]{} Rau A., Schady P., Greiner J. et al., 2012, A\&A, 538, A26

\bibitem[]{} Sbarrato T., Padovani P. \& Ghisellini G., 2014, MNRAS, 445, 81

\bibitem[]{} Sharma P., Quataert E., Hammett G.W., Stone J.M., 2007, ApJ, 667, 714 

\bibitem[]{} Sikora M., Begelman M.C. \& Rees M.J., 1994, ApJ, 421, 153 

\bibitem[]{} Taylor J.R., 1997, An introduction to error analysis, Univ. Science books 

\bibitem[]{} Urry, C.M. \& Padovani P., 1995, PASP, 107, 803

\bibitem[]{} Wall J.V. \& Peacock J.A., 1985, MNRAS, 216, 173



\end{thebibliography}
\end{document}